\documentclass[%
reprint,
superscriptaddress,
 amsmath,amssymb,
 aps,
]{revtex4-1}

\usepackage{graphicx}% Include figure files
\usepackage{dcolumn}% Align table columns on decimal point
\usepackage{bm}% bold math
\usepackage{threeparttable}
\usepackage{amssymb}
\usepackage{amsthm}
\usepackage{amsmath}
\usepackage{lineno}
\usepackage{hyperref}% add hypertext capabilities
\begin{document}

%\preprint{APS/PRA}

\title{Production of clean rare isotope beams at TRIUMF\\
Ion guide laser ion source}% Force line breaks with \\

\author{M. Mostamand}
\email{maryam@triumf.ca}
\affiliation{TRIUMF, 4004 Wesbrook Mall, Vancouver, BC, Canada, V6T 2A3}%
\affiliation{University of Manitoba, Winnipeg, MB, Canada,  R3T 2N2}%

\author{R. Li}
\affiliation{TRIUMF, 4004 Wesbrook Mall, Vancouver, BC, Canada, V6T 2A3}%

\author{J. Romans}
\affiliation{TRIUMF, 4004 Wesbrook Mall, Vancouver, BC, Canada, V6T 2A3}%
\affiliation{KU Leuven, Instituut voor Kern-en Stralingsfysica 200D, B-3001, Leuven, Belgium}%

\author{F. Ames}
\affiliation{TRIUMF, 4004 Wesbrook Mall, Vancouver, BC, Canada, V6T 2A3}%
\affiliation{Saint Mary’s University, Halifax, NS, Canada, B3H 3C3}%

\author{P. Kunz}
\affiliation{TRIUMF, 4004 Wesbrook Mall, Vancouver, BC, Canada, V6T 2A3}%
\affiliation{Simon Fraser University, Burnaby, BC, Canada, V5A 1S6}%

\author{A. Mj$\phi$s}
\affiliation{TRIUMF, 4004 Wesbrook Mall, Vancouver, BC, Canada, V6T 2A3}%

\author{J. Lassen}
\affiliation{TRIUMF, 4004 Wesbrook Mall, Vancouver, BC, Canada, V6T 2A3}%
\affiliation{University of Manitoba, Winnipeg, MB, Canada,  R3T 2N2}%
\affiliation{Simon Fraser University, Burnaby, BC, Canada, V5A 1S6}%

\date{\today}% It is always \today, today,

\begin{abstract}

Hot cavity resonant ionization laser ion sources (RILIS) provide a multitude of radioactive ion beams with high ionization efficiency and element selective ionization. However, in hot cavity RILIS there still remains isobaric contaminations in the extracted beam from surface ionized species.
An ion guide-laser ion source (IG-LIS) has been implemented that decouples the hot isotope production region from the laser ionization volume. A number of IG-LIS runs have been conducted to provide isobar free radioactive ion beams for experiments. Isobar suppression of up to 10$^{6}$ has been achieved, however, IG-LIS still suffers from an intensity loss of 50-100\,$\times$ as compared to hot cavity RILIS. 
Operating parameters for IG-LIS are being optimized and design improvements are being implemented into the prototype for robust and efficient on-line operation. Recent SIMION ion optics simulation results and the ongoing development status of the IG-LIS are presented.  

\end{abstract}


\maketitle

%\tableofcontents
\section{Introduction}
\label{intro}

\begin{figure}[!ht]
\centering
\includegraphics[width=\linewidth]{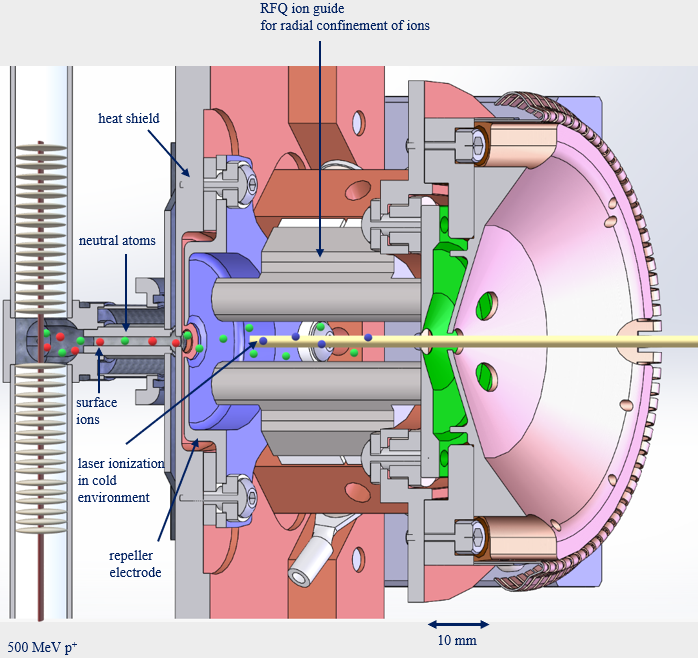}
\caption{Technical realization of the IG-LIS at ISAC. Shown is a cross section of the device, centered on the 3\,mm diameter transfer tube that connects to the top of the target container. It features from left to right: isotope production target, transfer tube, heat shield, repeller electrode, RFQ structure, exit electrode, and extraction electrode \cite{heggen_2013}. Potentials are chosen so that surface ionized species from the hot cavity/target transfer tube are prevented from entering the "cold" ionization volume within the ion guide.}
\label{iglis}
\end{figure}

Modern experiments at isotope separator on-line (ISOL) facilities like TRIUMF's isotope separator and accelerator (ISAC) often depend critically on the purity of the delivered rare isotope beams. Therefore, highly selective ion sources are required.
The mass separator magnet at ISAC was designed for $A/q \leq 30$ and an ion source emittance $\leq 30$\,$\pi$mm.mrad. As such the $m/\delta m \sim 2000$ with high throughput allows for isobar separation. As ISAC radioactive ion beams delivered have expanded to $A/q \leq 240$, there is need for additional selectivity.
Isobaric beam contamination often arises from surface ionization of elements with low ionization potential ($\leq 6$\,eV) on the hot target or transfer tube surfaces. The resulting background can be especially large for exotic nuclides far from stability and produced at low rates which stems from the fact that the abundance ratio becomes very unfavorable from a production point of view for 'isotope of interest / background isobar'. The electronic structure of elements, gives an avenue to use laser resonance ionization to provide element-selective ionization and enhance the yield of the isotope of interest. Although resonant laser ionization increases the yield of a desired element, delivered beams can still be overwhelmed by surface ionized isobars. In these cases, the production of a sufficiently pure laser ionized beam requires a means of improving the RILIS selectivity. This problem was tackled by developing a new type of ion source, the ion guide laser ion source (IG-LIS). This type of ion source was proposed by Blaum \textit{et al}. \cite{blaum_novel_2003} in the form of the laser ion source trap (LIST) as a laser resonance ionization within a segmented radio frequency quadrupole (RFQ). 
A first implementation of an IG-LIS (previously named RFQ-LIS) for off-line tests at ISAC was done in 2007 with a design \cite{lavoie_production_2010,lavoie_segmented_2007} for isobar suppression. Along the RFQ axis the electrodes are segmented into six sections. By applying DC potentials of $0-5$\,V to the individual segments a potential gradient can be created inside the RFQ to assist longitudinal guidance and extraction of the ions. Through detailed thermal and ion optics simulation studies and off-line tests with stable isotopes \cite{heggen_2013,raeder_ion_2014} a more robust and higher efficiency design without segmentation was brought on-line in 2013. This ion guide laser ion source is presented in Fig.~\ref{iglis}.
In a first on-line run with a silicon carbide (SiC) target a suppression of surface-ionized sodium (Na) contaminants in the ion beam of up to six orders of magnitude was demonstrated \cite{raeder_ion_2014}.
In this work, operation and further improvements of this unique laser ion source were tested, an IG-LIS will allow key experiments on the exotic ISOL beams.
A number of IG-LIS runs have been conducted at ISAC successfully.
Details about mechanical design and performance of the currently used on-line IG-LIS are given in \cite{heggen_2013}. 
This paper focuses on optimized operating parameters and design improvements for IG-LIS from SIMION \cite{dahl} simulations for a robust and efficient on-line operation. The energy spread of ions extracted from the ion-guide is investigated for different operation parameters. The influence of laser parameters on the mass separator transmission of extracted laser ionized species was calculated. A systematic study of the ion signal dependence on repetition rate and laser pulse energy has been performed in offline tests in Leuven \cite{rept_rate_2012,rept_rate_2013}.
The energy spread in an octupole ion-guide in comparison to the standard quadrupole was studied.

\section{Simulations and improvements on IG-LIS}
\label{sec:iglis}

The quadrupole ion guide based IG-LIS prototype operated at ISAC in 2013 aimed to guide ions and therefore operated in RF-only mode. It uses a square wave RF which allows simple adjustment of the driving frequency and duty cycle. With a field radius of r$_{0}$ = 5\,mm, a frequency range of 1--3\,MHz and RF-amplitudes up to 100\,V, singly charged ions with masses from 2 to 240$^{+}$\,amu can be transmitted.
IG-LIS design for on-line operation has to be radiation hard and compatible with high temperature, high vacuum conditions. The technical requirements and constraints are explained in details in \cite{heggen_2013,raeder_ion_2014}.
To develop a next generation IG-LIS, a better physical understanding of the IG-LIS prototype, ion optics, and its limiting factors is essential and can be arrived through simulations.
Further charged particle trajectory simulations of IG-LIS were carried out in SIMION 8 in view of different DC bias configurations and the use of higher multipole ion guides \cite{dahl}. 

\subsection{Potential array calculations along the beam axis}
\label{sec:potential}

\begin{figure}[!h]
\centering
\includegraphics[width=\linewidth]{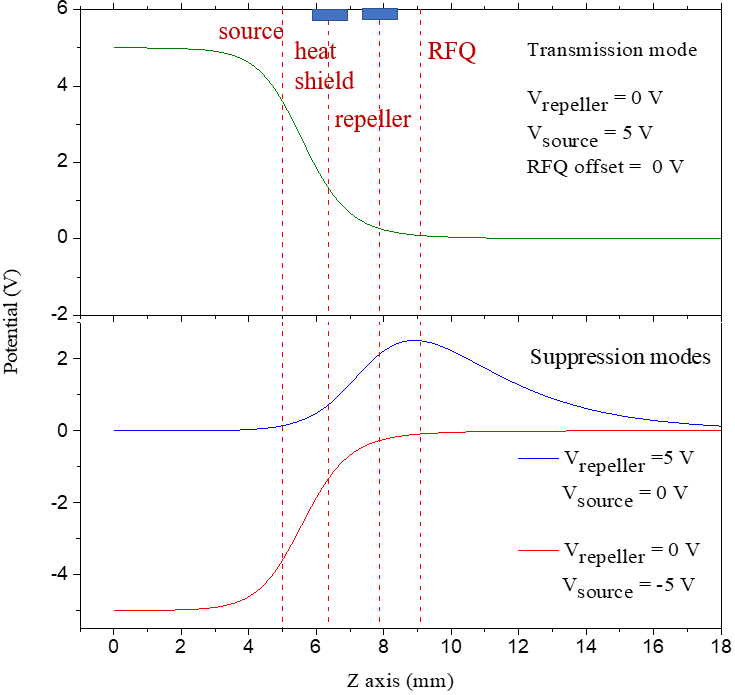}
\caption{Potential distribution along the z-axis in IG-LIS for the principal operation modes: (i) transmission and (ii) suppression mode either with (a) positively biased suppression electrode, or (b) negatively biased target. Vertical red dash lines represent the locations for source, heat shield, repeller electrode, and beginning of the 35\,mm long RFQ. V$_{repeller}$: potential on the repeller, V$_{source}$: potential on the source, RFQ offset: dc potential offset of RFQ. The region from which ions can be extracted is larger in transmission mode and allows for an improved ion extraction from the source without much surface ion suppression. The efficiency of the IG-LIS in transmission mode comes close to that achieved by the standard surface ion source - RILIS configuration. Source bias operation in suppression mode (red curve) compared to repeller operation (blue curve), covers a larger ionization region and has lower potential difference along the axis of cold ionization volume and therefore presents the highest efficiency IG-LIS operation mode. Blue rectangles on the top axis represent the thickness of the electrode. All potentials given are with respect to the source bias.}
\label{potentiala}
\end{figure}

The repeller electrode which is located 1.4\,mm behind the heat shield has a large aperture of 5\,mm diameter to avoid the heat radiation from the target and material deposition. The RFQ electrodes are 35\,mm long and are mounted on a single piece copper bracket. The electrodes are insulated from the copper mount by ceramic washers. This mount also holds an electrically insulated exit electrode with an aperture of 3.5\,mm diameter.
There are two different operation modes for IG-LIS: transmission mode, and suppression mode. Fig.~\ref{potentiala} presents the on axis potential in both operation modes. Positive source potential in the transmission mode helps to extract more ions from the source towards the RFQ and extraction electrode.

\begin{figure}[!h]
\centering
\includegraphics[width=\linewidth]{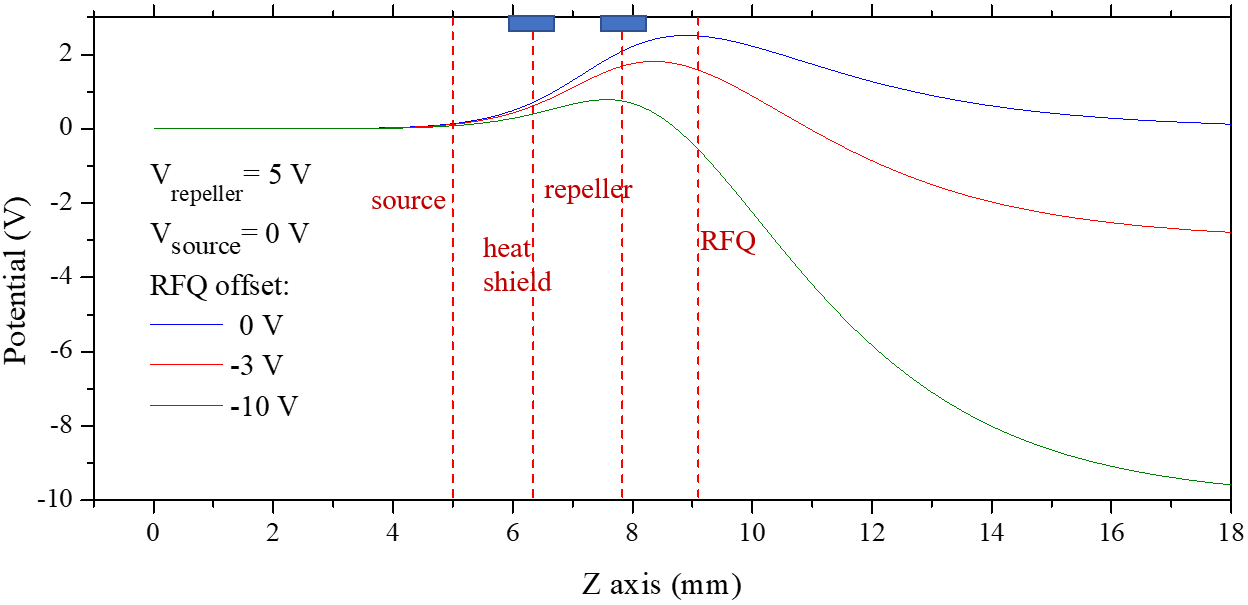}
\caption{Effect of RFQ bias operation with potential distribution along the z-axis in IG-LIS. Red dash lines represent the locations of source exit, heat shield, repeller electrode, and start of the RFQ. V$_{repeller}$: potential on the repeller, V$_{source}$: potential on the source, RFQ offset: dc potential offset of RFQ. Shifting down the potential on offset with respect to the repeller helps to extract more ions. Blue rectangles on the top axis represent the thickness of the electrode.}
\label{potential}
\end{figure}

By applying a voltage higher than the source bias to the repeller electrode, mainly neutral atoms will be able to enter the ionization region where they can be ionized by resonant laser light while surface ionized species are repelled. 
For online operation, beam tuning and optimization in high transmission mode is advantageous, since it provides intense beams for beam tuning.
Potential arrays along the z-axis in IG-LIS for the two operation modes with the standard on-line operation parameters are shown in Fig.~\ref{potential}.

The RFQ offset potential in Fig.~\ref{potential} helps to increase the kinetic energy of ions towards the RFQ and extraction electrode and consequently enhance the transmission efficiency (larger ionization region). However, the effect of this potential difference along the axis on the overall energy spread of the ions must be considered.
The mass separator's mass resolution of $m/\delta m \sim 2000$ has such an effect that ions with an energy spread above $\sim 10$\,eV will not be able to pass through the mass separator. Therefore the energy spread of ions extracted from the IG-LIS must be considered.

\subsection{Energy spread}
\label{sec:Es}

\begin{figure}[!h]
\centering
\includegraphics[width=\linewidth]{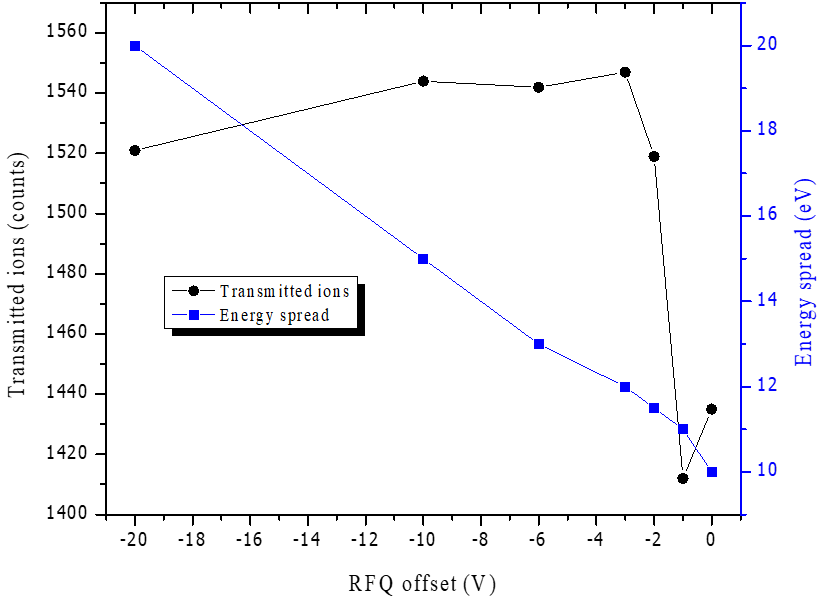}
\caption{Effect of RFQ-offset on extracted ion's energy spread and transmission efficiency in suppression mode. Transmission efficiency saturates at RFQ-offset = -3\,V.}
\label{data_offset}
\end{figure}

For on-line IG-LIS operation, the effect of energy spread on the extracted beam and transmission efficiency through the downstream mass separator is an important factor. The resultant simulation studies presented in this section are particularly useful for establishing high resolution, high throughput mass separator tunes for radioactive isotope beams from IG-LIS.
Enhancement of the beam transmission (in suppression mode) to some extent with shifting down the offset potential is reasonable, since lowering the RFQ offset with respect to the source causes larger kinetic energy of the ions and therefore pushes more ions from the source towards the extraction electrode. However, shifting the offset potential results in a larger potential difference in the ionization region and consequently higher energy spread. 
Fig.~\ref{data_offset} presents the effect of RFQ-offset on ion transmission and energy spread in suppression mode. This effect saturates at RFQ-offset = -3\,V after which there is no further increase in transmission efficiency.

\section{Possible IG-LIS operational and design improvements}
\label{sec:simulations}

Simulation studies in this section provide possible improvements of energy spread and overall efficiency in the ion guide.

\subsection{Effect of laser parameters on the ion guide efficiency}
\label{sec:laser rep}

\begin{figure}[!h]
\centering
\includegraphics[width=\linewidth]{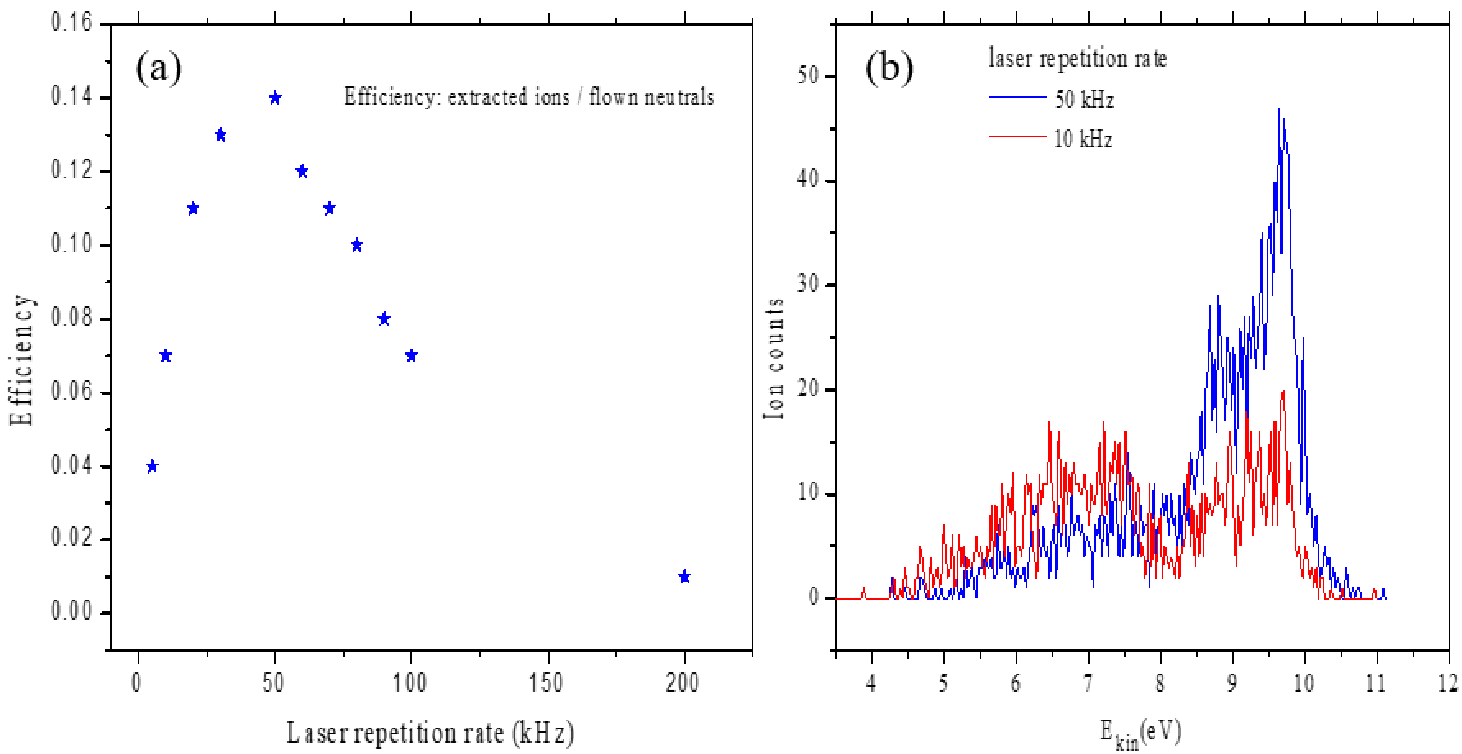}
\caption{Calculated laser repetition rate effect on (a) the extracted laser ionized species and (b) energy spread of the extracted laser ionized beam. IG-LIS ionization reaches optimum efficiency at a repetition rate of 50\,kHz. Whether this maximum point is exactly at 50\,kHz or not depends on the ionization efficiency of the laser scheme which is considered to be 100\% for simulations in this work. Simulation is performed on mass 200\,amu.}
\label{laserep}
\end{figure}

Calulated IG-LIS transmission efficiency as a function of laser repetition rate is shown in Fig.~\ref{laserep}.

An increased laser repetition rate helps to ionize more of the neutral atoms that enter the ion guide region, resulting in higher efficiency. 
Calculations are based on suppression mode operation and mass 200\,amu.
With increasing duty cycle eventually the low energy tail in Fig.~\ref{laserep}(red curve in (b)) can be suppressed and most of the ions are extracted with the same energy.
At repetition rates higher than 50\,kHz in suppression mode, most of the neutral atoms are laser ionized before they pass the repeller and IG-LIS efficiency is reduced. Thus most of the laser ionized species will be rejected by the repeller electrode and the efficiency drops down. 
An experimental test of the laser repetition rate behavior was not possible due the the performance characteristics of the available pump lasers (LEE LDP100MQg) which are limited in overall output power and repetition rate maximum of 20\,kHz. A 5 fold increase in repetition rate would also require the equivalent increase in pump power from currently $\sim$10\,W/Ti:Sa laser to effectively 50\,W/Ti:Sa laser. The optimized operating conditions at on-line ISOL type facilities are based on pulsed laser operation to make use of the high pulse peak power for hot cavity RILIS the optimum repetition rate is at about 10\,kHz, which fits well with available laser technology. In the standard hot cavity RILIS atoms are confined in the approximately 40\,mm long, 3\,mm diameter transfer tube, with much larger residence time, as well as a re-introduction of atoms into the ionization volume through spatial confinement of atoms such that ion yields are optimal at $\sim$ 10\,kHz laser repetition rate.
\subsection{Octupole ion guide}
\label{sec:octupole}

In an octupole RF ion-guide the concentration of the field near the field radius is more pronounced.
The octupole field is usually approximated using eight cylindrical rods. 
An octupole RF ion guide can transport a range of masses four times larger than the quadrupole ion guide of the same field radius and field parameters. Due to concentration of the field close to the field radius the amplitude of particle motion in an octupole guide is larger than in the quadrupole with similar field parameters. From the study of octupole potential distributions for several geometries, each with different rod-to-field-radius ratios an optimum value of r$_{rod}$/r$_{0}$ = 0.355 was derived \cite{niculae_numerical_2009}.
The kinetic energies of ions are modulated much more strongly by the quadrupole RF field than higher multipoles RF field. The ion beam from the ion source has an intrinsic energy, position, and angular spread (phase space in mm.mrad or emittance), any defect in the primary beam would affect beam resolution significantly, due to compression and distortion by the multi-keV axial acceleration and focusing abberations prior to inserting into the magnet mass separator system.   
The transverse kinetic energy of an ion propagating inside a linear RF ion guide does not stay constant. It varies rapidly following the ion's rapid micro-motion and the slower macro-motion in the RF field in the radial plane. For higher order multipoles, the RF field in the center is much weaker, and the micro-motion is less pronounced near the axis.
For a better acceptance and resolution by the mass separator magnet (ISAC at TRIUMF) with $\delta E \sim 1$\,eV, more ions can be extracted from an octupole ion guide.

\subsubsection{Octupole versus quadrupole ion guide}

\begin{figure}[!ht]
\centering
\includegraphics[width=\linewidth]{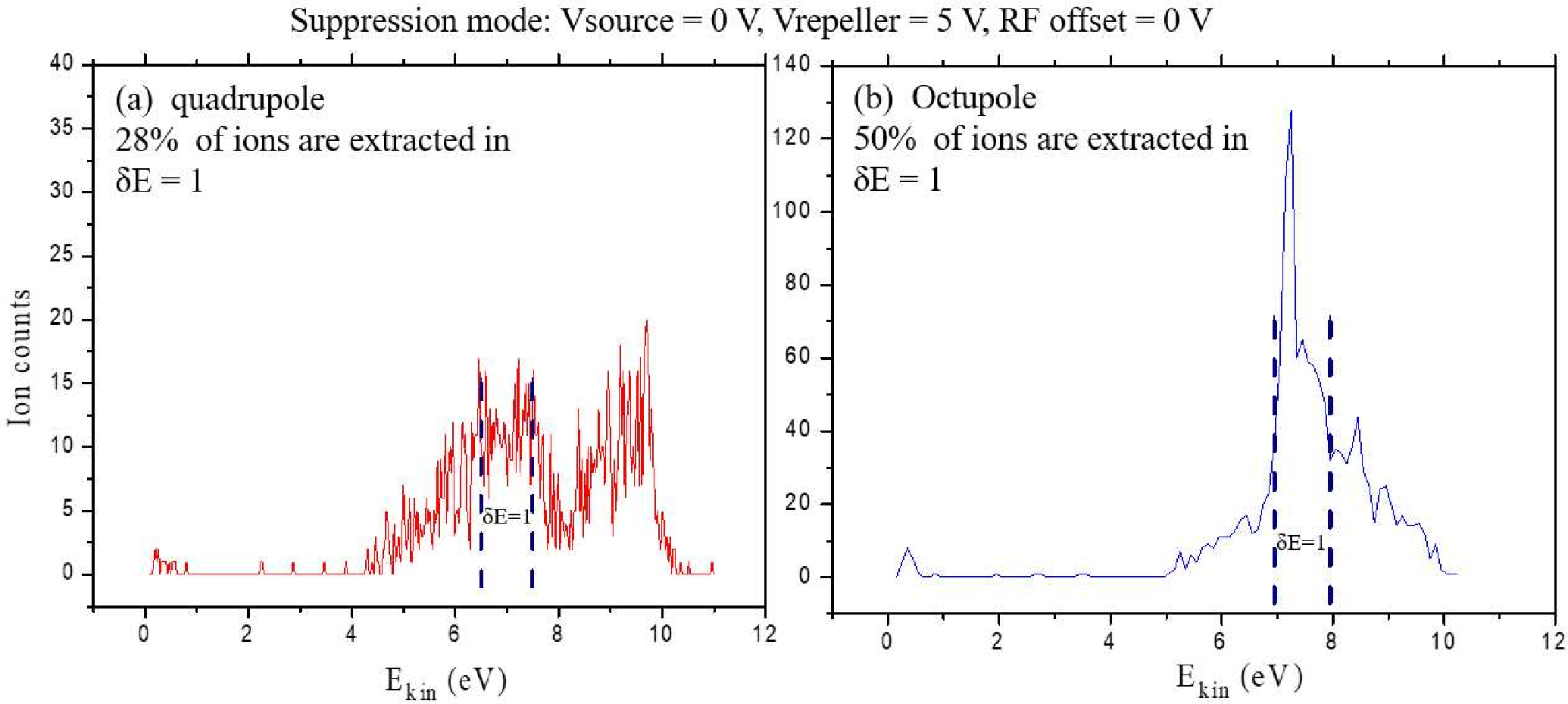}
\caption{Simulation studies for kinetic energy of extracted ions in suppression mode from (a) a quadrupole ion guide and (b) an octupole ion guide. Dashed lines represent the $\delta E$ = 1\,eV region for extracted ions from an ion guide. Due to smaller radial RF field gradient of an octupole, more of the ions are extracted within 1\,eV transmission window. Simulations are performed for mass 200\,amu.}
\label{quadoct}
\end{figure}

Simulation studies depicted in Fig.~\ref{quadoct}(a) show that the energy spread of ions extracted from a quadrupole ion-guide configuration only allows to extract a maximum 28\% of ions at the magnet mass separator with an energy acceptance of $\sim$\,1\,eV. Upgrading to an octupole ion guide, Fig.~\ref{quadoct}(b), would increase the efficiency to 50\% with the same 1\,eV energy spread.

\begin{figure}[!ht]
\centering
\includegraphics[width=\linewidth]{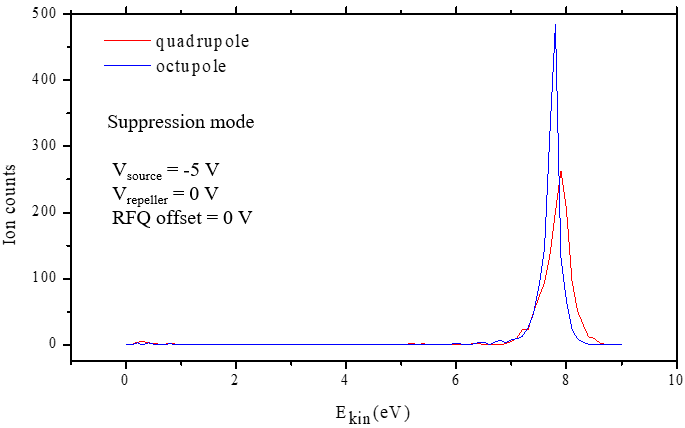}
\caption{Simulation studies for kinetic energy of extracted ions in suppression mode with source bias (the preferred IGLIS operation mode) from a quadrupole ion guide (red curve) and an octupole ion guide (blue curve). The peak in blue curve indicates a reduced energy spread of the extracted ions in an octupole ion guide. Simulations are performed for mass 200\,amu.}
\label{bias}
\end{figure}

An alternative method to suppress surface ions is to bias the source instead of applying a potential on the repeller electrode. 
Utilizing a source-bias setup results in a reduced ion energy spread over 1\,eV compared to the suppression by means of the repeller electrode for both quardupole and octupole ion guides (Fig.~\ref{bias}). It allows for a more efficient transmission of ions through the mass separator.

\section{Conclusion}
 
Operating the IG-LIS (RFQ-offset, RF-amplitude and frequency) with proper settings is critical in order for system performance.
Modifying the IG-LIS ion guide to an octupole results in a net reduction of energy spread. Low energy spread becomes important when coupling IG-LIS to the ISAC isobar separator for collinear fast beam laser spectroscopy, or when injecting beams into ion traps. Increasing laser repetition rate from 10\,kHz to 50\,kHz can enhance the IG-LIS efficiency while significantly reducing the energy spread of extracted laser ions.

\section*{Acknowledgments}

The work has been funded by TRIUMF under a contribution from National Research Council of Canada (NRC) and through a Natural Sciences and Engineering Research Council of Canada (NSERC) Discovery Grant (SAP-IN-2017-00039). M. Mostamand acknowledges funding through the University of Manitoba Graduate Fellowship.\\
\\


\begin{thebibliography}{}


\bibitem{blaum_novel_2003}
K. Blaum, C. Geppert, H. J. Kluge, M. Mukherjee, S. Schwarz and K. Wendt, A novel scheme for a highly selective laser ion source, Nucl. Instrum. Meth. B. 204, 331--335 (2003).

\bibitem{lavoie_production_2010}
J. P. Lavoie, Production of pure ion beams by laser ionization and a fast release {RFQ}, PhD thesis, University of Laval (2010).

\bibitem{lavoie_segmented_2007}
J. P. Lavoie, P. Bricault, J. Lassen and M. R. Pearson, Segmented linear radiofrequency quadrupole/laser ion source project at {TRIUMF}, Hyperfine Interact. 174, 33--39 (2007).

\bibitem{heggen_2013}
H. Heggen, Development of a radio frequency quadrupole- laser ion source (RFQ - LIS) for isobar suppression, M.Sc thesis, Technische Universit\"at Darmstadt (2013).

\bibitem{raeder_ion_2014}
S. Raeder, H. Heggen, J. Lassen, F. Ames, D. Bishop, P. Bricault, P. Kunz, A. Mj$\phi$s and A. Teigelh\"ofer, An ion guide laser ion source for isobar-suppressed rare isotope beams, Rev. Sci. Instrum. 85, 033309 (2014).

\bibitem{dahl}
D. A. Dahl, SIMION for the personal computer in reflection, Int. J. Mass. Spect. 200, (2000).

\bibitem{niculae_numerical_2009}
C. Niculae, and M. Niculae, Numerical method for calculating of potential distribution in non-ideal multipole ion guides, Optoelectron. Adv. Mater. Rapid. Commun. 3, 1073--1075 (2009).

\bibitem{rept_rate_2012}
R. Ferrer, V. T. Sonnenschein, B. Bastin, S. Franchoo and M. Huyse, Performance of a high repetition pulse rate laser system for in-gas-jet laser ionization studies with the Leuven laser ion source @ LISOL, Nucl. Instrum. Meth. B. 291, 29--37 (2012).

\bibitem{rept_rate_2013}
T. Kron, R. Ferrer, N. Lecesne, V. Sonnenschein, S. Raeder, J. Rossnagel and K. Wendt, Control of RILIS lasers at IGISOL facilities using a compact atomic beam reference cell, Hyperfine Interact. 216, 53--58 (2013).

\end{thebibliography}
\end{document}